\documentclass[sigconf,nonacm]{acmart}

\setcopyright{none}
\usepackage{algorithm}
\usepackage[noend]{algpseudocode}
\usepackage{subcaption} 
\usepackage{tikz}
\usepackage{tcolorbox}
\usepackage{tabularx}
\usepackage{placeins} 

\newlength{\RoundedBoxWidth}
\newsavebox{\GrayRoundedBox}
   {\setlength{\RoundedBoxWidth}{\dimexpr#1}
    \begin{lrbox}{\GrayRoundedBox}
       \begin{minipage}{\RoundedBoxWidth}}%
   {   \end{minipage}
    \end{lrbox}
    \begin{center}
    \begin{tikzpicture}%
       \draw node[draw=black,fill=black!10,rounded corners,%
             inner sep=2ex,text width=\RoundedBoxWidth]%
             {\usebox{\GrayRoundedBox}};
    \end{tikzpicture}
    \end{center}}

\newcolumntype{Y}{>{\raggedleft\arraybackslash}X}

\begin{document}

\title{Lead-Lag Relationships in Financial Markets: A Comparison of Multiple Clustering Algorithms}

\author{Ruichen Deng}
\affiliation{%
  \institution{Nankai University}
  \country{China}
}
\email{2311274@nankai.edu.cn}

\author{Yichi Zhang}
\affiliation{%
  \institution{University of Oxford}
  \country{UK}
}
\email{yichi.zhang@stats.ox.ac.uk}

\begin{abstract}
Lead-lag relationships are widely used in financial time series, and many clustering algorithms based on them have been developed. The traditional DTW-KMedoids algorithm performs well both on the synthetic dataset and the real financial dataset. However, there are still several limitations to these algorithms: low efficiency caused by high time complexity, poor mathematical properties from DTW distance, the clustering effect is sensitive to the number of clusters. To solve the problems above and improve the performance, this paper introduces three clustering algorithms: MiniRocket-KMeans, KShape, Ensemble algorithm (a combination of KShape and DTW-KMedoids) and compares their performance on synthetic and real stock datasets with DTW-KMedoids algorithm under the same trade strategy. In addition, this paper also finds the best number of clusters by maximizing the silhouette coefficient in each clustering algorithm to improve the stability of the experiment results. Our main conclusions are as follows: MiniRocket-KMeans performs best under the lead strategy, achieving a Sharpe ratio of 0.866 with a maximum drawdown controlled at -63.9\%; the ensemble algorithm exhibits excellent stability; the robustness is significantly improved after finding the best number of clusters; the p-values of the hypothesis test on the Sharpe ratio of all strategies are 0.0, verifying the statistical validity of the lead-lag trading strategy. Finally, future improvement directions such as customized lead-lag matrices and optimized ensemble voting mechanisms are proposed.
\end{abstract}

\keywords{Financial time series analysis, Dynamic time warping, MiniRocket-KMeans, KShape, DTW-KMedoids, Ensemble clustering, Lead-lag relationships}

\maketitle

\section{Introduction}

In recent years, with the development of machine learning theory and practice, a large number of researchers have attempted to apply it to financial time series to explore the potential structures of such high-dimensional time series. For example, lead-lag relationships are particularly common and significant in the financial field \cite{DTW_leadlag_lagged, ito2020direct, albers2021fragmentation, buccheri2021highfrequency, zhang2023robust}. Traditional methods for mining lead-lag relationships using cross-correlation suffer from insufficient consistency and high time complexity \cite{multiref_leadlag}. Zhang et al. (2023) \cite{zhang2023robust} proposed using DTW to quantify lead-lag relationships between asset pairs, complemented by the RowSum ranking method to delineate leading groups (Leaders) and lagging groups (Laggers). Notably, this integrated approach demonstrates strong robustness and consistency in empirical validations.

DTW is an algorithm used to quantify the similarity between two time series. It can accurately capture similar patterns despite temporal misalignment through an optimal alignment path \cite{bellman1959adaptive, berndt1994dynamic, keogh2005exact, DTW_review} and is widely applied to explore lead-lag relationships between asset pairs. However, its high time complexity results in low efficiency when clustering large-scale stock data. Furthermore, the distance metric defined by DTW does not satisfy the triangle inequality, which may compromise clustering accuracy \cite{DTW_review}.

In fact, apart from the aforementioned algorithm, there are many clustering algorithms capable of detecting lead-lag relationships.

MiniRocket is a feature extraction algorithm specifically designed for large-scale time series. By reconstructing the logic of the original Rocket model, it achieves a maximum speedup of 75 times while maintaining nearly consistent classification accuracy and possesses an almost deterministic nature, realizing the dual breakthrough of high precision and high speed. It can complete the training and testing of 109 datasets in the UCR time series archive within 10 minutes, providing an efficient solution for large-scale time series classification \cite{minirocket}. Using it as the underlying algorithm can significantly improve computational efficiency while retaining accuracy.

KShape is an efficient shape-based clustering algorithm. Through innovative shape distance measurement and cluster centroid calculation methods, the algorithm achieves far superior efficiency and accuracy compared to traditional algorithms while maintaining invariance to amplitude scaling and phase shifts, making it one of the benchmark algorithms in the field of time series clustering \cite{KShape}. Compared with DTW-based algorithm, its advantages lie in robustness and computational efficiency.

This paper introduces three different clustering algorithms: 

MiniRocket-KMeans, KShape, the ensemble algorithm (a combination of DTW-KMedoids and KShape) and compares their performance with DTW-KMedoids under the same trading strategy. Also, we find the best number of clusters by maximizing the silhouette coefficient to improve the stability of the experiment results.

\textbf{Our study is organized around the following four main contributions and findings:}
\begin{enumerate}
\item After using the number of clusters with maximal silhouette coefficient, the stability of cluster is significantly improved and better performance is also achieved in the final backtesting.
\item The hard voting ensemble model combining the traditional DTW-KMedoids algorithm and KShape exhibits exceptional stability. However, its drawback is that it is overly conservative under the trading strategy, wasting some effective lead-lag pairs. It is suitable for scenarios with high noise, large sample sizes and a focus on stability. 
\item The MiniRocket-KMeans algorithm outperforms the DTW-KMedoids algorithm in backtesting under the lead-lag trading strategy, featuring a smaller maximum drawdown and higher cumulative returns. Meanwhile, KShape demonstrates significant advantages in terms of the risk-reward ratio.
\item These three new algorithms outperform the DTW-KMedoids algorithm in different aspects, providing more options for various scenarios. Additionally, after determining the optimal cluster number automatically, the adaptability of the DTW-KMedoids algorithm is improved, enabling more robust clustering results with different quantities and types of stocks.
\end{enumerate}

\section{Theoretical Introduction to the Algorithms}

\subsection{DTW-KMedoids Algorithm}

Dynamic Time Warping (DTW) measures time-series similarity by finding an optimal alignment path that minimizes warping cost, effectively handling temporal misalignment and phase shifts \cite{bellman1959adaptive, berndt1994dynamic}. KMedoids is subsequently applied to achieve partition. Unlike KMeans, its cluster centroids are chosen from actual existing sample points (medoids) rather than virtual mean points, which yields stronger robustness against financial market noise and allows integration with the non-Euclidean DTW distance matrix \cite{DTW_leadlag_lagged}.

\subsection{KShape Algorithm}

KShape is an iterative, shape-based clustering method \cite{KShape}. It defines the shape-based distance (SBD) derived from normalized cross-correlation (NCC) of two series $X$ and $Y$:
\begin{equation}
\operatorname{SBD}(X, Y)=1-\max _{\tau} \frac{(X \star Y)[\tau]}{\|X\|\|Y\|},
\end{equation}
where $\star$ denotes cross-correlation and $\tau$ denotes the time lag. Under the assignment-update iterative framework, cluster centroids are updated to maximize cross-correlation within each cluster. KShape maintains invariance to scaling and phase shifts while avoiding expensive pairwise DTW calculations.

\subsection{Ensemble Algorithm}

To mitigate single-algorithm classification bias, the ensemble method applies hard voting to combine DTW-KMedoids and KShape. Let $L_{\mathrm{DTW}}$ and $L_{\text{KShape}}$ be independent sets of clustering labels. An asset pair is grouped into the same cluster if and only if both underlying algorithms assign them to the same cluster. This approach prioritizes classification stability and drastically reduces noise-induced misclassifications under volatile market conditions.

\subsection{MiniRocket-KMeans Algorithm}

MiniRocket is an ultra-fast feature extraction framework that generates temporal features using lightweight convolution kernels \cite{minirocket}. It applies dilated convolutions to capture local patterns (trends, peaks, and valleys) and maps each series $X$ to a feature vector in Euclidean space using the positive part value (PPV):
\begin{equation}
\operatorname{PPV}\left(C_{i}(X)\right)=\frac{1}{T-l_{i} \cdot d_{i}+1} \sum_{t=1}^{T-l_{i} \cdot d_{i}+1} \mathbb{I}\left\{C_{i}(X)[t]>0\right\},
\end{equation}
where $l_i$ is kernel length, $d_i$ is dilation, and $\mathbb{I}\{\cdot\}$ is the indicator function. Traditional KMeans is then performed on the generated PPV feature vectors. This reduces clustering time for large asset portfolios from hours to minutes.

\subsection{Determine the Best Cluster Number $K$}

To automate cluster selection, we optimize $K$ using the silhouette coefficient:
\begin{equation}
s(i)=\frac{b(i)-a(i)}{\max (a(i), b(i))},
\end{equation}
where $a(i)$ is the average intra-cluster distance and $b(i)$ is the average distance to the nearest neighboring cluster. The optimal $K$ is selected by maximizing the average silhouette coefficient $S = \frac{1}{n} \sum s(i)$, which enhances model adaptability across varying portfolio sizes.

\subsection{Leaders and Laggers}

Within each cluster, assets are partitioned into Leaders (signals) and Laggers (targets) using the RowSum ranking method \cite{DTW_leadlag_lagged}.

\subsubsection{Construction of the Lead-Lag Matrix}
For assets $i$ and $j$, DTW alignments yield local lag values $\delta_{ij}^k$. These are aggregated into a global lag $\hat{L}_{ij}$ using either mode estimation (robust to noise) or median estimation (capturing overall distribution):

\begin{itemize}

\item Mode estimation: Take the value with the highest frequency among the local lag values, which has strong robustness to sporadic abnormal offsets caused by noise, its formula is as follows:
\begin{equation}\hat{L}_{i j}^{mod}=\arg \max _{v} \operatorname{count}\left(\delta_{i j}^{k}=v\right), \quad k=1, \ldots, K.\end{equation}

\item Median estimation: Take the middle value of the sorted local lag values, which reflects the overall distribution center of lag values but is susceptible to continuous offsets, its formula is as follows:
\begin{equation}\hat{L}_{i j}^{med}=\operatorname{median}\left(\left\{\delta_{i j}^{1}, \ldots, \delta_{i j}^{K}\right\}\right).\end{equation}

\end{itemize}

The symmetric lead-lag matrix $\mathbf{M}$ is defined as:
\begin{equation}
\mathbf{M}[i, j]=\hat{L}_{i j}-\hat{L}_{j i}.
\end{equation}

\subsubsection{Leading and Lagger Partition}
Row-wise summing yields the global leading score $S_{i}=\sum_{j} \mathbf{M}[i, j]$. Sorting these scores in descending order, the top $\alpha$ (0.25) fraction of assets with the lowest scores (most leading) form the Leader set $G_L$, while the remaining form the Lagger set $G_D$.
\section{Synthetic Data Experiments}

In this section, synthetic data are used to test the clustering performance and anti-interference ability of several models.

\subsection{Synthetic Data Generation}

We generate data using a lagged multi-factor model based on the standard multi-factor model in multivariate time series. This setup effectively simulates the characteristics that stock returns are driven by multi-dimensional factors and different stocks exhibit time lags in their responses to factors, making the data closer to real market data. The theoretical foundation is as follows:
\begin{itemize}
\item The value \(X_{i}^{t}\) of time series \(i\) at time \(t\) under the standard multi-factor model is expressed as:
\begin{equation}X_{i}^{t}=\sum_{j=1}^{k} B_{i j} f_{j}^{t}+\epsilon_{i}^{t} \quad i=1, \ldots, n ; \quad t=1, \ldots, T,\end{equation}
where \(k\) is the number of factors, \(B_{i j}\) is the exposure of time series \(i\) to factor \(j\), \(f_{j}^{t}\) is the value of factor \(j\) at time \(t\), \(\epsilon_{i}^{t}\) is the noise term at time \(t\) with variance \(\sigma^2\), \(n\) is the total number of time series and \(T\) is the total number of time steps.
\item Its lagged version is expressed as:
\begin{equation}X_{i}^{t}=\sum_{j=1}^{k} B_{i j} f_{j}^{t-L_{i j}}+\epsilon_{i}^{t} \quad i=1, \ldots, n ; \quad t=1, \ldots, T.\end{equation}
The difference from the standard multi-factor model is the introduction of \(L_{i j}\), which denotes the lag order corresponding to the exposure of time series \(i\) to factor \(j\). Thus, \(f_{j}^{t-L_{i j}}\) is the value of factor \(j\) at time \(t-L_{i j}\).
\end{itemize}

To simplify the model, this paper assumes that each time series has lagged exposure to only one single factor. In fact, this setting was proposed in \cite{DTW_leadlag_lagged} and is referred to as \textbf{Single Membership}. In contrast, \textbf{Mixed Membership} allows each time series to have lagged exposure to multiple factors, forming a mixed structure.

In addition, there are two sub-settings under Single Membership:
\begin{itemize}
\item \textbf{Homogeneous Setting}: The model contains only one factor, i.e., \(K=1\).
\item \textbf{Heterogeneous Setting}: The model contains more than one factor (\(K\ge 2\)), but each time series has exposure to only one single factor.
\end{itemize}

\subsection{Performance Comparison on Synthetic Data}

To simplify the experiments, we only conduct tests on the most representative and complex case: the heterogeneous setting with \(K=3\). The conclusions can be effectively generalized.

\subsubsection{Comparison of ARI and MSE Between Different Algorithms}

\begin{figure*}[t] %
    \centering
    \begin{subfigure}{0.32\textwidth} %
        \centering
        \includegraphics[width=\linewidth]{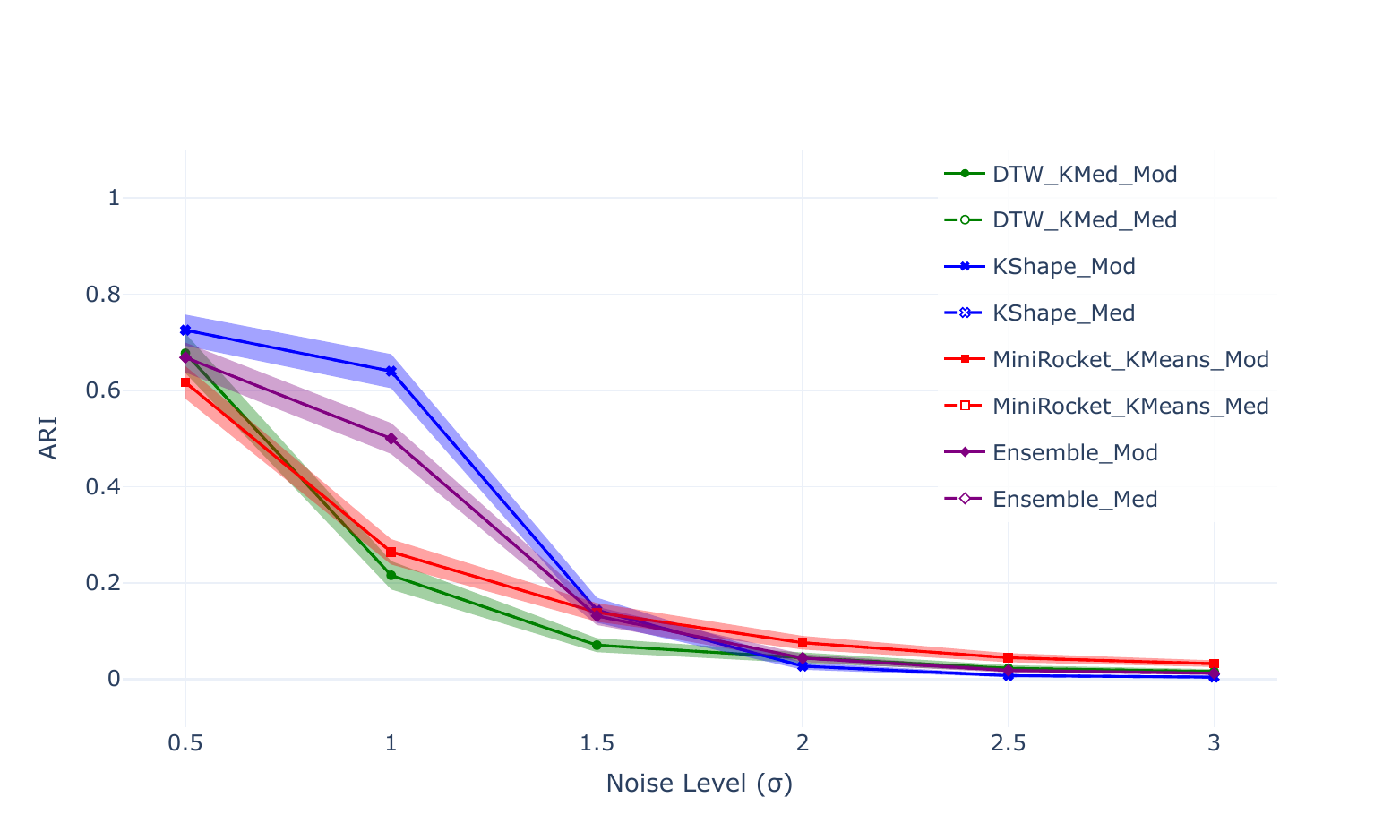}
        \caption{Varying noise levels ($\sigma$)}
        \label{fig:ari_noise}
    \end{subfigure}
    \hfill
    \begin{subfigure}{0.32\textwidth}
        \centering
        \includegraphics[width=\linewidth]{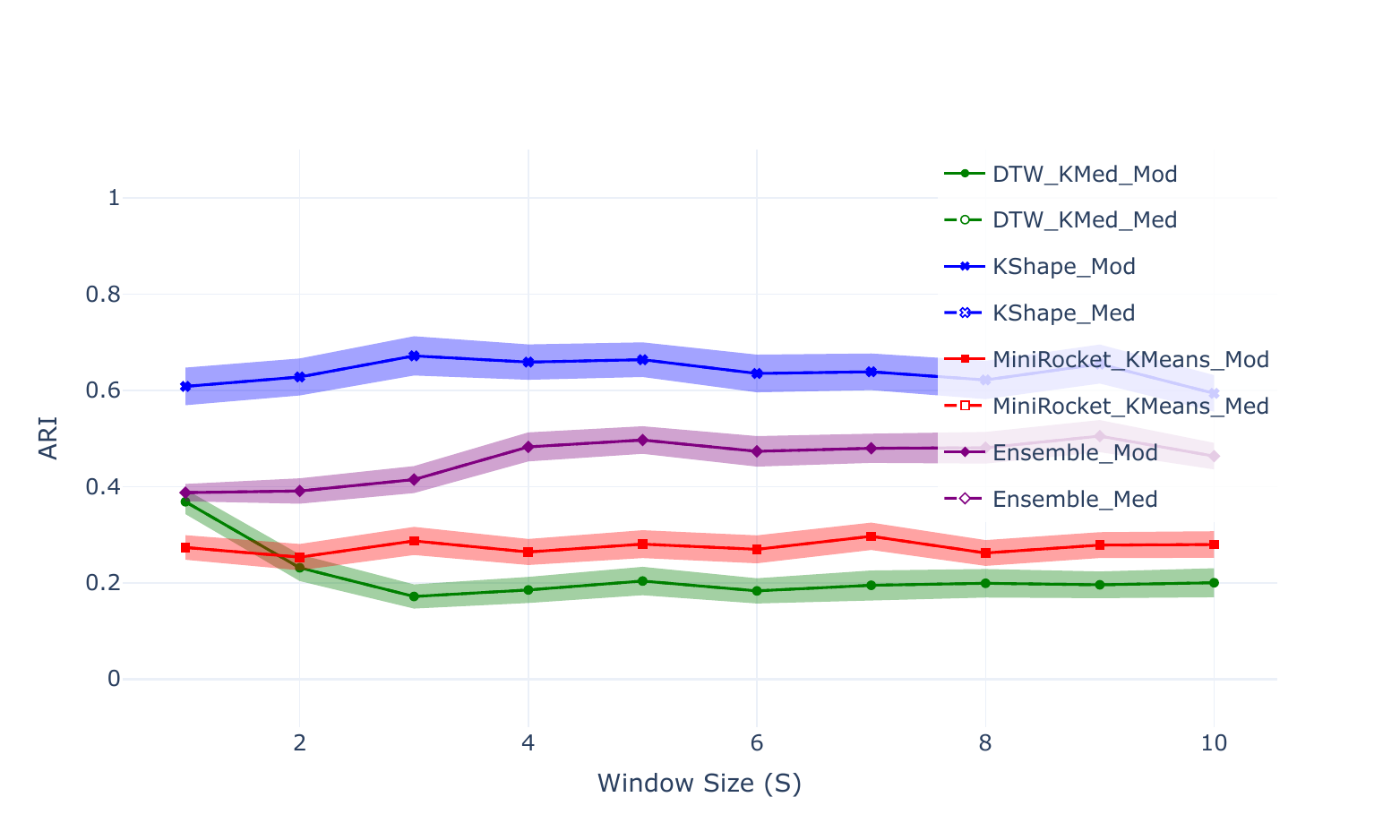}
        \caption{Small window sizes}
        \label{fig:ari_win_small}
    \end{subfigure}
    \hfill
    \begin{subfigure}{0.32\textwidth}
        \centering
        \includegraphics[width=\linewidth]{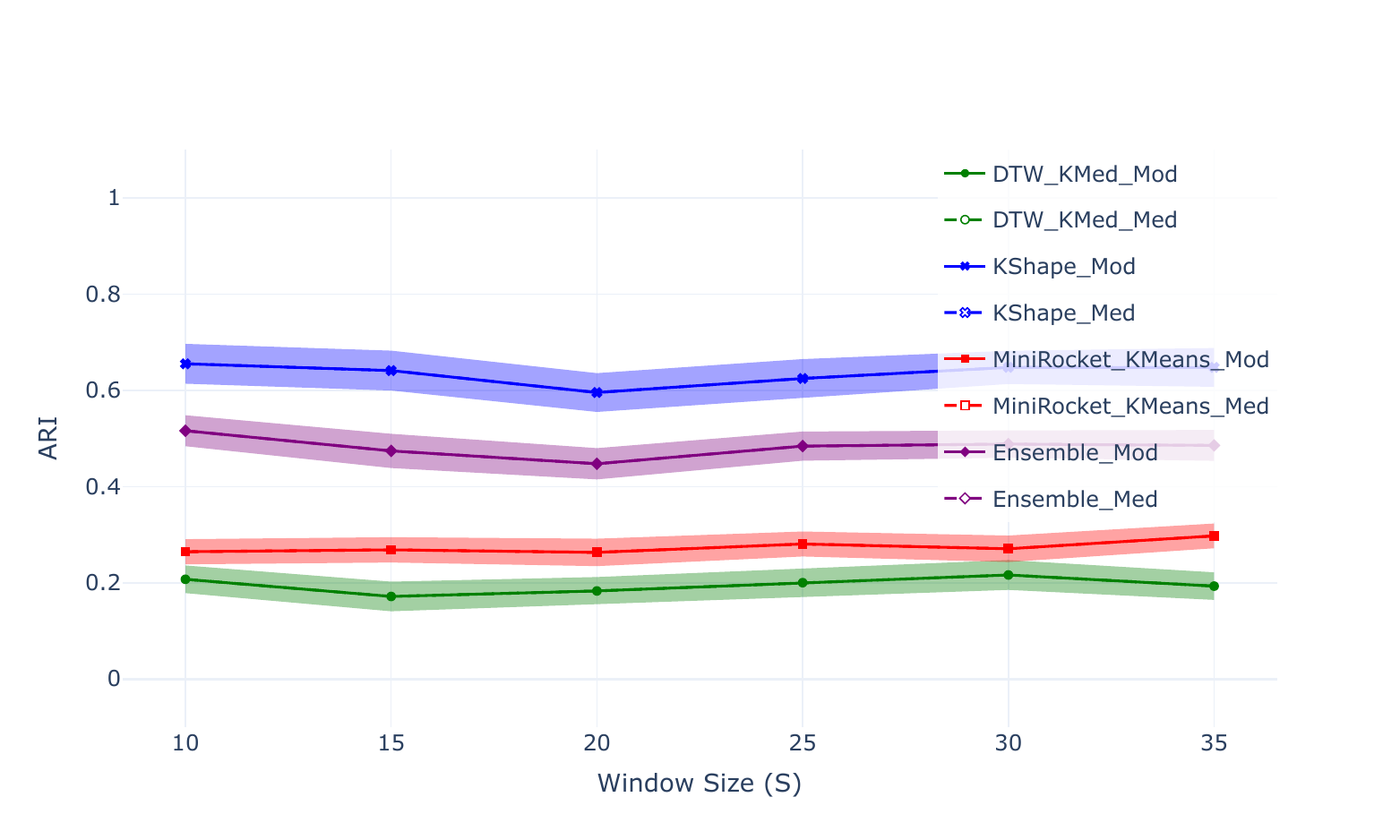}
        \caption{Large window sizes}
        \label{fig:ari_win_large}
    \end{subfigure}
    \caption{Clustering ARI comparison under different noise levels and window sizes.}
    \label{fig:ari_group}
\end{figure*}

In this section, we use ARI and MSE as evaluation metrics for clustering performance.

\textbf{Theoretical Introduction to ARI and MSE:}
ARI is a metric that measures the consistency between clustering results and true labels. It addresses the bias of the original rand index (RI) toward random clustering and is computed based on a contingency table. Let \(a\) be the number of sample pairs that belong to the same true class and the same cluster; \(b\) is the number of sample pairs that belong to different true classes and different clusters; \(n\) is the total number of samples; the total number of sample pairs is $N = \binom{n}{2}$. The adjusted rand index (ARI) is defined as:
\begin{equation}ARI=\frac{RI-E[RI]}{\max(RI)-E[RI]}.\end{equation}
ARI ranges in \([-1,1]\). A value closer to \(1\) indicates higher consistency. 

MSE measures the deviation between estimated and true values. In this paper, it is used to evaluate the error between the estimated lag matrix and the true lag matrix, with the formula:
\begin{equation}MSE=\frac{1}{n \times n} \sum_{i=1}^{n} \sum_{j=1}^{n}\left(\hat{L}_{i j}-L_{i j}\right)^{2},\end{equation}
where \(\hat{L}_{ij}\) is the estimated lag of series \(i\) relative to series \(j\), and \(L_{ij}\) is the true lag.

\textbf{Experimental Procedure and Results:}
In the experimental procedure, the lagged multi-factor model ($K=3$) is used to generate synthetic data. We investigate the effects of noise level and window size on clustering performance by adjusting these parameters. Each experiment is repeated 100 times to obtain the mean values and 95\% confidence intervals.


\begin{figure*}[t]
    \centering
    \begin{subfigure}{0.32\textwidth}
        \centering
        \includegraphics[width=\linewidth]{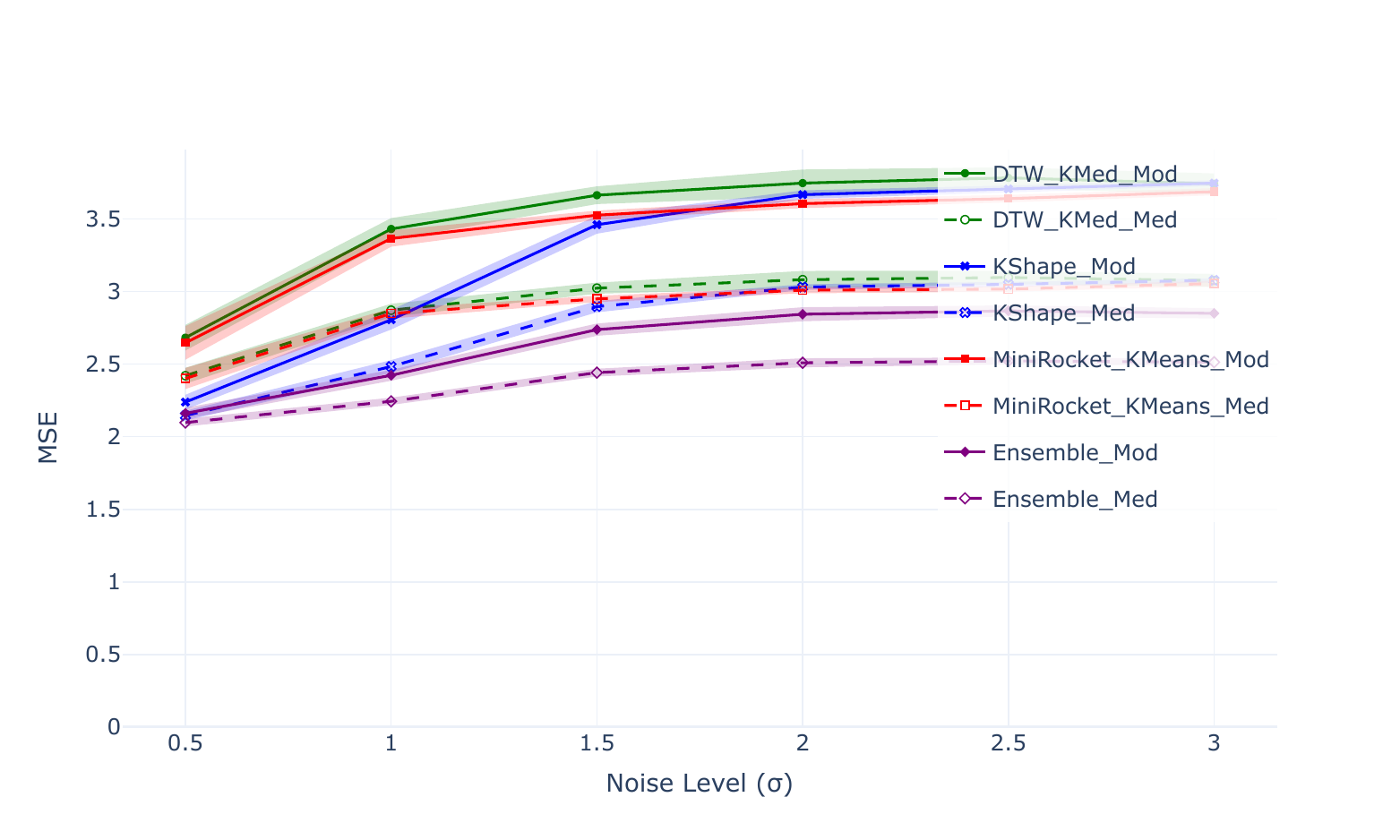}
        \caption{Varying noise levels ($\sigma$)}
        \label{fig:mse_noise}
    \end{subfigure}
    \hfill
    \begin{subfigure}{0.32\textwidth}
        \centering
        \includegraphics[width=\linewidth]{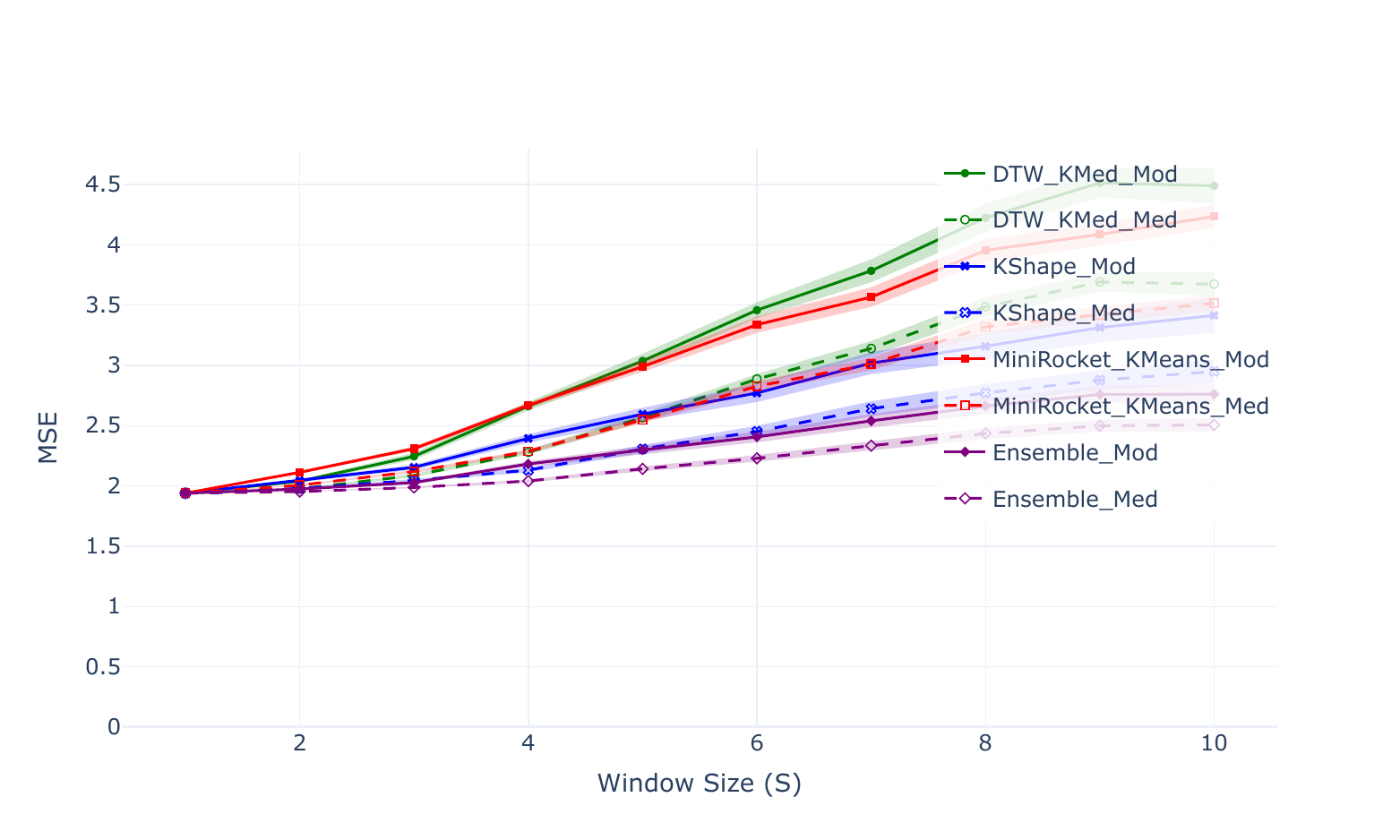}
        \caption{Small window sizes}
        \label{fig:mse_win_small}
    \end{subfigure}
    \hfill
    \begin{subfigure}{0.32\textwidth}
        \centering
        \includegraphics[width=\linewidth]{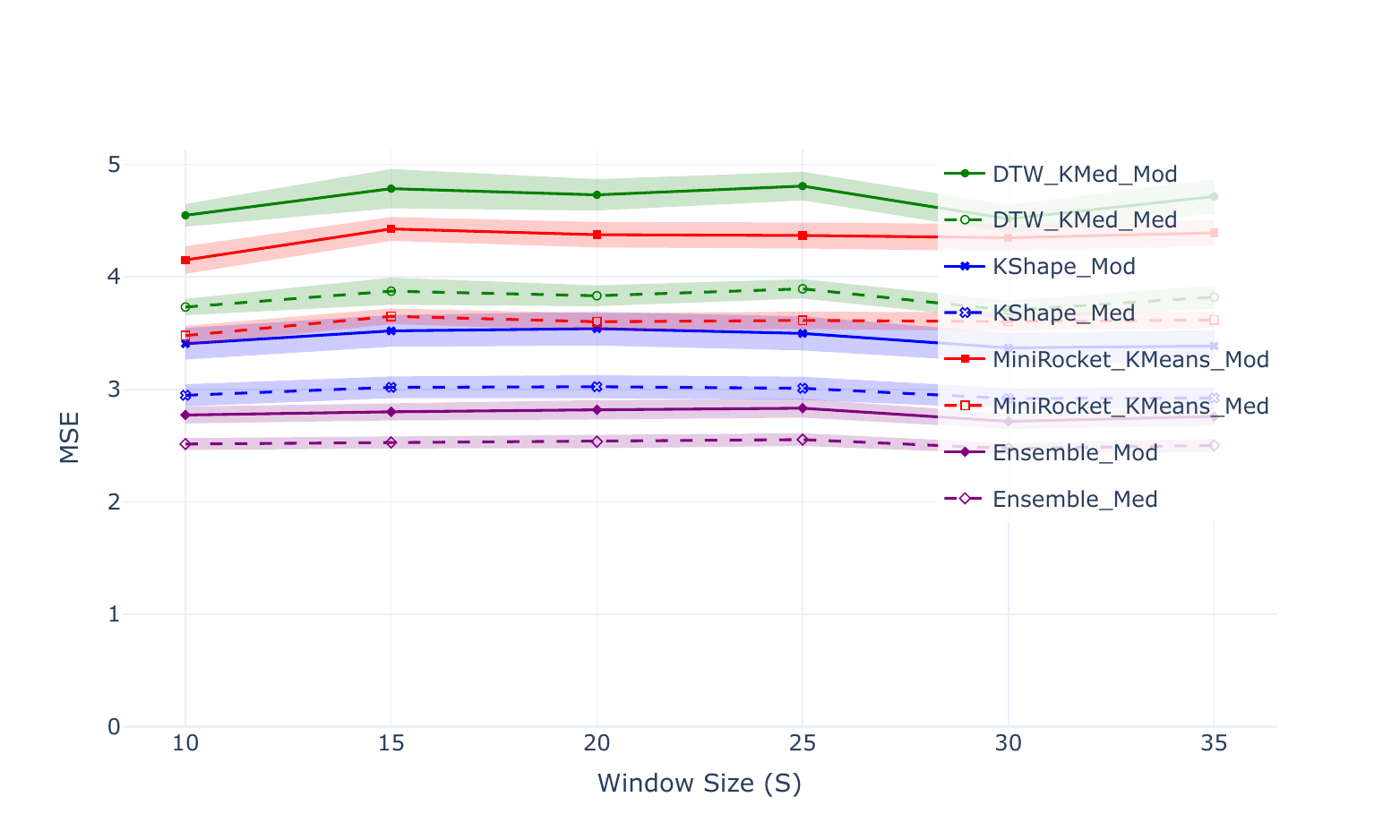}
        \caption{Large window sizes}
        \label{fig:mse_win_large}
    \end{subfigure}
    \caption{Lag matrix MSE comparison under different noise levels and window sizes.}
    \label{fig:mse_group}
\end{figure*}

\begin{itemize}
    \item In terms of ARI (shown in Figure \ref{fig:ari_group}), it can be observed that as the noise level $\sigma$ increases from 0.5 to 3.0, the ARI of all algorithms shows a significant decreasing trend. Under low noise conditions, the other three algorithms overall outperform the DTW-KMedoids algorithm, among which KShape achieves the best clustering performance. Under high noise conditions, the MiniRocket-KMeans algorithm yields the optimal results. After determining the optimal cluster number automatically, the ARI of the four algorithms does not change significantly across different window sizes. Under different window sizes, the ARI values of the other three algorithms are all higher than that of the DTW-KMedoids algorithm.
\end{itemize}

Note: Here, med and mod correspond to the two methods (median / mode) for calculating the lead-lag values between asset pairs when constructing the lead-lag matrix. Since the lead-lag matrix is not computed during the clustering process itself, the curves for the two methods coincide exactly when calculating ARI. However, since MSE measures the error between the estimated lag matrix and the true lag matrix, the curves under the med and mod methods do not overlap.


\begin{itemize}
    \item In terms of MSE (shown in Figure \ref{fig:mse_group}), it can be observed that the MSE of all algorithms shows an upward trend as the noise level increases. Overall, the MSE values obtained using the median are generally lower than those using the mode, indicating that the median provides a more accurate estimation for computing the lead-lag matrix. The ensemble algorithm performs the best and the other two algorithms also outperform the DTW-KMedoids algorithm. When the window size is below 10, the MSE of all algorithms increases with the window size. When the window size exceeds 10, the MSE of each algorithm does not fluctuate significantly. The MSE values of the three algorithms are also generally lower than that of the DTW-KMedoids algorithm, with the ensemble algorithm still achieving the best performance.
\end{itemize}

In summary, we have identified algorithms that outperform the DTW-KMedoids algorithm on synthetic data under different scenarios and metrics. 

\section{Real Data Experiments}

Financial markets represent a typical scenario where lead-lag relationships naturally exist. In this section, we conduct large-scale experiments using real financial data. Two datasets of daily closing prices for different stocks are used, with details summarized in Table \ref{tab: dataset}.

\begin{table*}[t]
  \centering
  \caption{Summary of the two financial data sets considered in the experiments.}
  \label{tab: dataset}
  \footnotesize
  \setlength{\tabcolsep}{8pt}
  \begin{tabular}{lcccccc}
        \toprule
            \textbf{Data source} & \textbf{Type} & \textbf{Freq} & \textbf{\#\ of assets} & \textbf{Start date} & \textbf{End date} & \textbf{\#\ of days}\\
        \midrule
            Wharton’s CRSP     & Equity  & Daily & 679 & 2000/01/03 & 2019/12/31 & 5211\\
            Wharton’s CRSP     & Equity     & Daily & 1028  & 2000/01/31 & 2019/07/01 & 4880\\
        \bottomrule
  \end{tabular}
\end{table*}

\subsection{Comparison of Clustering Performance Under Different Numbers of Stocks}

\begin{figure*}[t]  
    \centering  
    \begin{subfigure}{0.45\textwidth}
        \centering  
        \includegraphics[width=\linewidth]{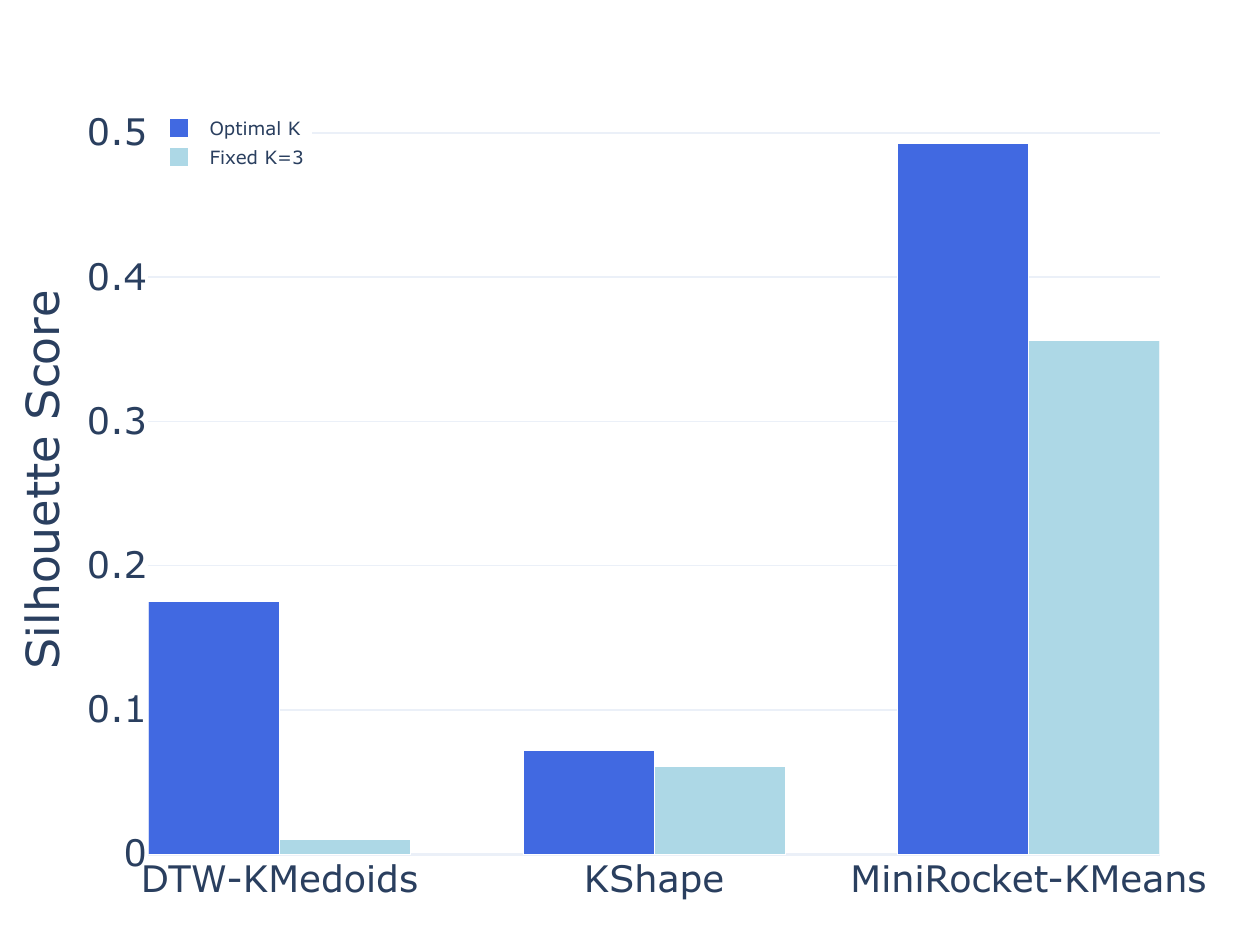}  
        \caption{Silhouette comparison on 679 assets.}  
        \label{fig:compa600}  
    \end{subfigure}
    \hfill  
    \begin{subfigure}{0.45\textwidth}
        \centering  
        \includegraphics[width=\linewidth]{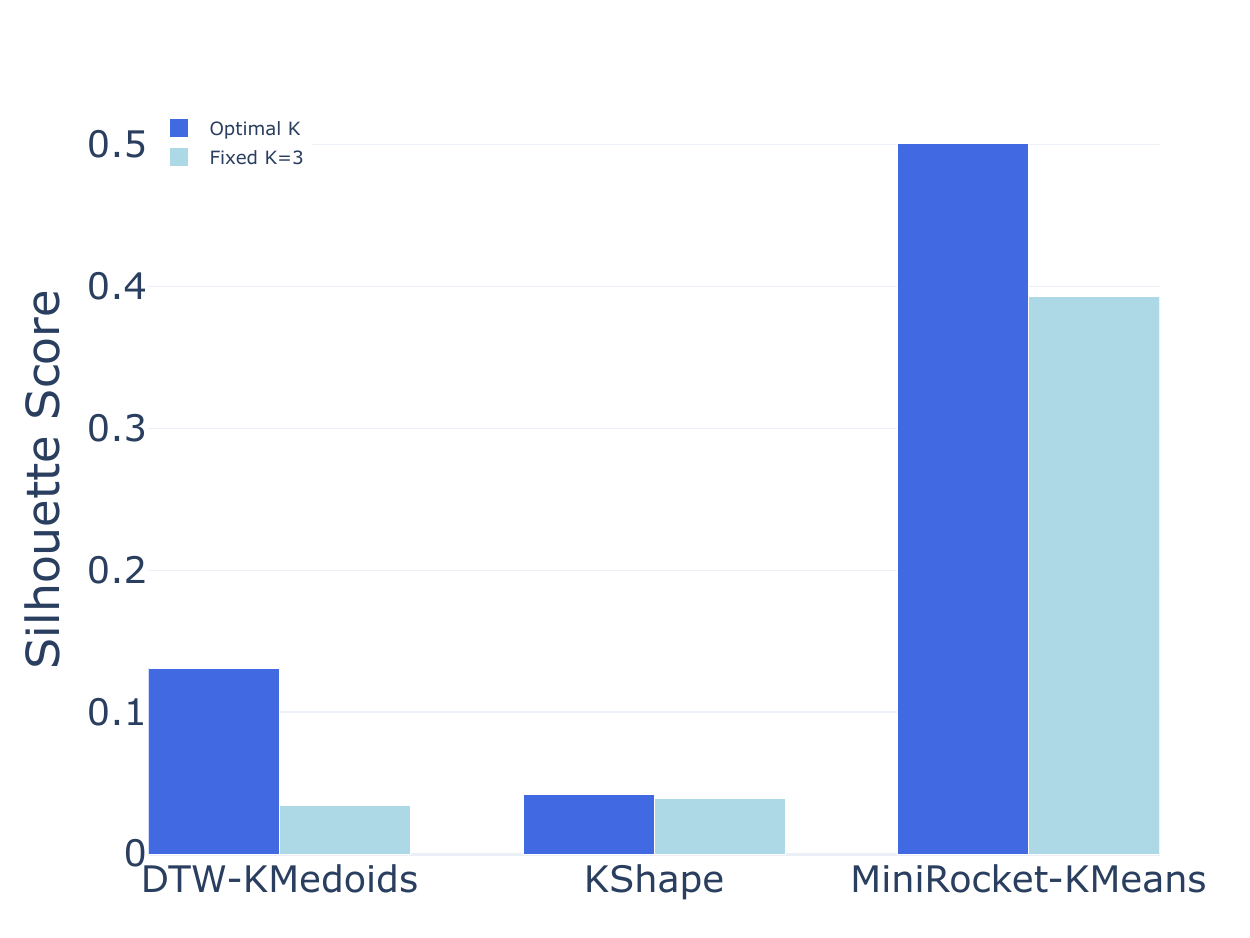}  
        \caption{Silhouette comparison on 1028 assets.}  
        \label{fig:compa1000}  
    \end{subfigure}
    \caption{Silhouette comparison under optimal and fixed cluster numbers across different datasets.}
    \label{fig:silhouette_comparison}
\end{figure*}

We first test the performance of the optimal cluster number selection algorithm using real data. The two datasets are clustered using different algorithms with the optimal number of clusters and the average silhouette coefficient of all samples is computed and compared with a fixed cluster number \(K=3\) (Figure \ref{fig:silhouette_comparison}).

It can be seen that with the optimal number of clusters, the performance of all algorithms is improved on both datasets.

\subsection{Trading Strategy}

The trading strategy is constructed based on time series clustering and lead-lag relationships, trading lagging assets using signals from leading assets. The specific steps are as shown in Algorithm \ref{strategy}.

\begin{algorithm}[htp] 
\caption{Trading Strategy}
\label{strategy}
\textbf{Input:} Time series matrix $X_{n \times T}$.
\begin{algorithmic}[1]
\State Extract sub-time series using a sliding window of fixed length $l=21$, obtaining $X_{n \times l}$ for each window.
\State Cluster the sub-time series within each window using the four proposed clustering algorithms.
\State For asset pairs within each cluster, apply the DTW-based lead-lag detection algorithm to divide them into Leader $\mathcal{L}$ and Lagger $\mathcal{G}$.
\State Compute the trading signal as the sign of the EWMA (span $spn$) of the average returns of the Leader:
\[
\text{signal} = \text{sign}\left( \frac{1}{|\mathcal{L}|} \sum_{k \in \mathcal{L}} \text{EWMA}(R_k, spn) \right)
\]
\State Calculate the daily PnL series based on the strategy type:
\begin{itemize}
\item If strategy type is \textbf{lead}:
\[
PnL_i = \text{signal} \cdot \frac{1}{|\mathcal{L}|} \sum_{k \in \mathcal{L}} R_k(t + i \cdot w + f)
\]
\item If strategy type is \textbf{lag}:
\[
PnL_i = \text{signal} \cdot \frac{1}{|\mathcal{G}|} \sum_{k \in \mathcal{G}} R_k(t + i \cdot w + f)
\]
\end{itemize}
\State Shift the sliding window by step $w$, and repeat Steps 1-5 until the end of the time series $X_{n \times T}$.
\end{algorithmic}
\end{algorithm}

We use momentum \cite{jegadeesh2001profitability, lim2019enhancing, poh2022transfer, tan2023spatiotemporal, wood2021trading, wood2021slow, roberts2023network, zohren2023learning} inside this framework. We take the top 75\% of time series after ranking as the Leader and the remaining as the Lagger. To forecast future performance, we use the exponentially weighted moving average (EWMA) of the average excess returns of the Leader over the past \(p = \{1, 3, 5, 7\}\) days to generate trading signals.

\subsection{Backtesting Metrics}

The backtesting metrics quantify strategy performance across four dimensions: profitability (annualized return $R_{\text{annual}}$), risk (annualized volatility $\sigma_{\text{annual}}$, maximum drawdown $MDD$), risk-adjusted return (Sharpe ratio $SR$), stability (hit/win rate $HR$, profit-loss ratio $PLR$), and statistical significance (Sharpe ratio hypothesis test $P$-value).

\subsection{Backtesting Results}

The performance metrics of the two datasets under various algorithm and trading strategies are presented in Table \ref{tab:trading_metrics} and Table \ref{tab:trading_metrics_1000}.

\begin{table*}[t]
  \centering
  \caption{Trading metrics for the dataset containing 679 assets.}
  \label{tab:trading_metrics}
  \footnotesize
  \renewcommand{\arraystretch}{1.22} 
  \setlength{\tabcolsep}{5.5pt}
  \begin{tabular}{lcccccccc}
  \toprule
    algorithm & strategy & Sharpe Ratio & Hit Rate & Max Drawdown & Profit Loss Ratio & Annual Return & Annual Volatility & Sharpe P Value \\
  \midrule
    DTW\_KMedoids\_mod & lag & 0.706 & 0.517 & -122.703 & 1.065 & 0.0512 & 0.0726 & 0.0 \\
    DTW\_KMedoids\_med & lag & 0.793 & 0.521 & -109.148 & 1.063 & 0.0583 & 0.0735 & 0.0 \\
    KShape\_mod & lag & 0.650 & 0.516 & -133.997 & 1.053 & 0.0492 & 0.0757 & 0.0 \\
    KShape\_med & lag & 0.579 & 0.514 & -125.684 & 1.049 & 0.0442 & 0.0763 & 0.0 \\
    MiniRocket\_KMeans\_mod & lag & 0.689 & 0.517 & -114.118 & 1.058 & 0.0507 & 0.0736 & 0.0 \\
    MiniRocket\_KMeans\_med & lag & 0.739 & 0.521 & -144.229 & 1.052 & 0.0548 & 0.0741 & 0.0 \\
    Ensemble\_mod & lag & 0.573 & 0.511 & -135.930 & 1.061 & 0.0410 & 0.0716 & 0.0 \\
    Ensemble\_med & lag & 0.609 & 0.518 & -132.002 & 1.040 & 0.0441 & 0.0725 & 0.0 \\
    DTW\_KMedoids\_mod & lead & 0.715 & 0.517 & -71.518 & 1.064 & 0.0521 & 0.0729 & 0.0 \\
    DTW\_KMedoids\_med & lead & 0.801 & 0.521 & -70.276 & 1.061 & 0.0580 & 0.0725 & 0.0 \\
    KShape\_mod & lead & 0.808 & 0.518 & -67.604 & 1.078 & 0.0575 & 0.0712 & 0.0 \\
    KShape\_med & lead & 0.790 & 0.515 & -69.418 & 1.089 & 0.0560 & 0.0708 & 0.0 \\
    MiniRocket\_KMeans\_mod & lead & 0.769 & 0.521 & -67.066 & 1.058 & 0.0554 & 0.0720 & 0.0 \\
    MiniRocket\_KMeans\_med & lead & \textbf{0.866} & 0.520 & -63.908 & 1.079 & 0.0621 & 0.0717 & 0.0 \\
    Ensemble\_mod & lead & 0.643 & 0.515 & -67.094 & 1.060 & 0.0465 & 0.0724 & 0.0 \\
    Ensemble\_med & lead & 0.679 & 0.519 & -67.879 & 1.050 & 0.0489 & 0.0720 & 0.0 \\
  \bottomrule
  \end{tabular}
\end{table*}

\begin{table*}[t]
  \centering
  \caption{Trading metrics for the dataset containing 1028 assets.}
  \label{tab:trading_metrics_1000}
  \footnotesize
  \renewcommand{\arraystretch}{1.22} 
  \setlength{\tabcolsep}{5.5pt}
  \begin{tabular}{lcccccccc}
  \toprule
    algorithm & strategy & Sharpe Ratio & Hit Rate & Max Drawdown & Profit Loss Ratio & Annual Return & Annual Volatility & Sharpe P Value  \\
  \midrule
    DTW\_KMedoids\_mod & lag & 0.371 & 0.502 & -843.835 & 1.061 & 0.0300 & 0.0808 & 0.0 \\
    DTW\_KMedoids\_med & lag & 0.353 & 0.500 & -400.600 & 1.065 & 0.0288 & 0.0816 & 0.0 \\
    KShape\_mod & lag & 0.321 & 0.515 & -259.902 & 0.998 & 0.0269 & 0.0836 & 0.0 \\
    KShape\_med & lag & 0.373 & 0.510 & -142.985 & 1.024 & 0.0314 & 0.0842 & 0.0 \\
    MiniRocket\_KMeans\_mod & lag & 0.405 & 0.510 & -1161.520 & 1.033 & 0.0336 & 0.0830 & 0.0 \\
    MiniRocket\_KMeans\_med & lag & 0.307 & 0.502 & -635.786 & 1.048 & 0.0256 & 0.0835 & 0.0 \\
    Ensemble\_mod & lag & \textbf{0.474} & 0.509 & -751.317 & 1.048 & 0.0385 & 0.0813 & 0.0 \\
    Ensemble\_med & lag & 0.447 & 0.508 & -451.613 & 1.047 & 0.0366 & 0.0820 & 0.0 \\
    DTW\_KMedoids\_mod & lead & 0.156 & 0.505 & -566.770 & 1.008 & 0.0132 & 0.0852 & 0.0 \\
    DTW\_KMedoids\_med & lead & 0.188 & 0.505 & -735.014 & 1.013 & 0.0160 & 0.0849 & 0.0 \\
    KShape\_mod & lead & 0.189 & 0.507 & -1766.753 & 1.004 & 0.0159 & 0.0840 & 0.0 \\
    KShape\_med & lead & 0.267 & 0.510 & -1029.415 & 1.008 & 0.0224 & 0.0838 & 0.0 \\
    MiniRocket\_KMeans\_mod & lead & 0.220 & 0.509 & -1071.332 & 1.002 & 0.0185 & 0.0838 & 0.0 \\
    MiniRocket\_KMeans\_med & lead & 0.160 & 0.505 & -629.837 & 1.006 & 0.0133 & 0.0836 & 0.0 \\
    Ensemble\_mod & lead & 0.251 & 0.509 & -471.088 & 1.008 & 0.0213 & 0.0848 & 0.0 \\
    Ensemble\_med & lead & 0.227 & 0.505 & -542.783 & 1.017 & 0.0192 & 0.0846 & 0.0 \\
  \bottomrule
  \end{tabular}
\end{table*}

The p-values of the hypothesis test on the Sharpe ratio of all algorithms are 0.0, indicating that the risk-adjusted returns of the strategies are statistically significant and not the result of random fluctuations. 

In most algorithms, the median estimator slightly outperforms the mode estimator. This is particularly evident in the lead strategy for the dataset containing 679 assets, where the Sharpe ratios of DTW\_KMedoids\_med, KShape\_med and MiniRocket\_KMeans\_med are all higher than their corresponding mod versions. 

MiniRocket-KMeans performs prominently under the lead strategy. For the lead strategy in the dataset containing 679 assets, the MiniRocket\_KMeans\_med algorithm achieves the highest Sharpe ratio (0.866) and annualized return (6.21\%), while controlling the maximum drawdown at -63.9\%, showing excellent risk-adjusted returns. In comparison, DTW-KMedoids and KShape algorithms, although stable, are slightly inferior in the combined performance of returns and drawdowns.

In terms of robustness, the ensemble algorithm stands out, mainly because it performs consistently across both datasets without extremely poor outcomes. It even achieves the best performance under the lag strategy in the dataset containing 679 assets. 





\begin{figure*}[t]  
\centering  
\begin{subfigure}{0.48\textwidth}  
    \centering  
    \includegraphics[width=\linewidth]{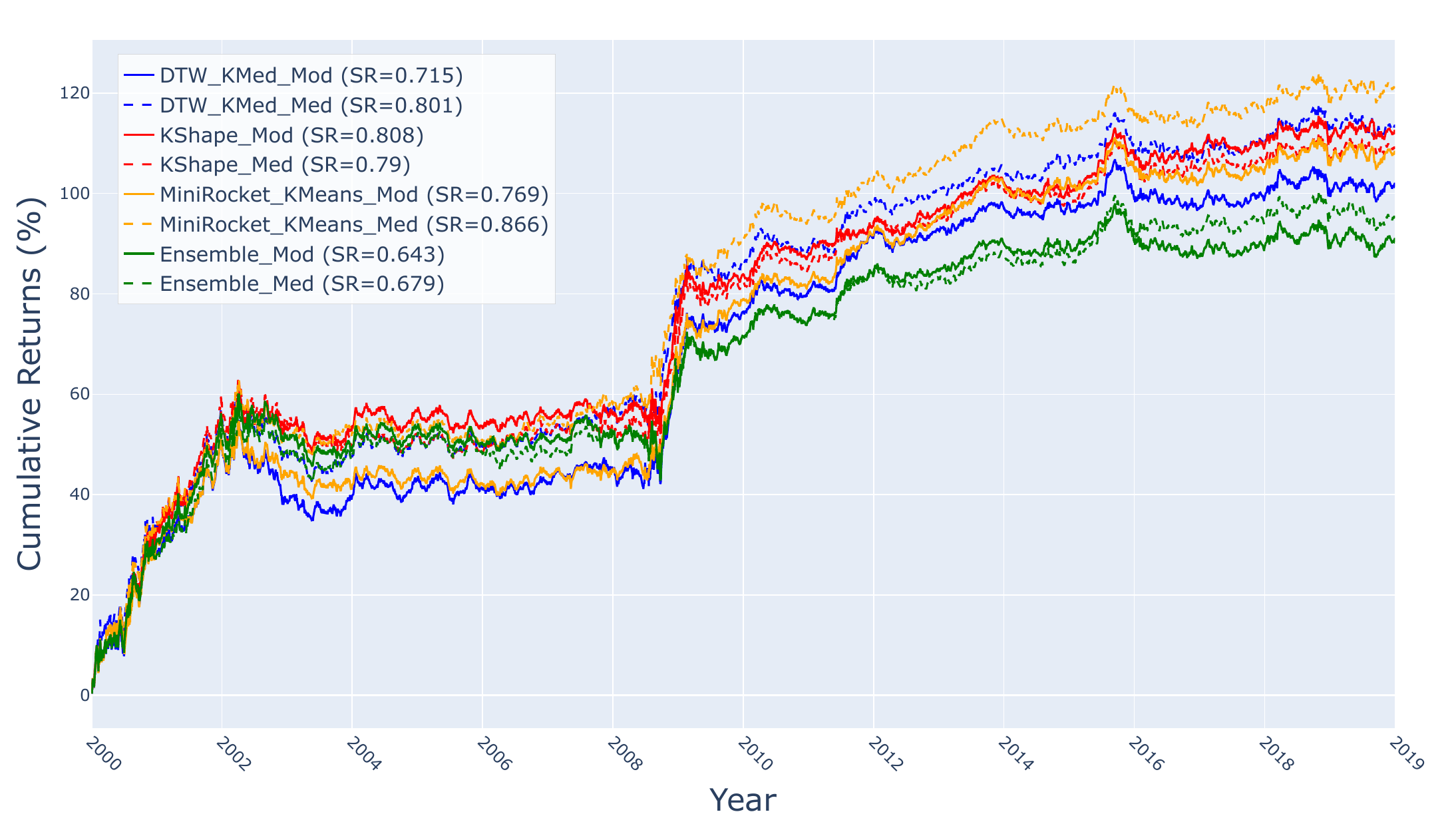}  
    \caption{Lead strategy on 679 assets.} 
    \label{fig:Nation}  
\end{subfigure}  
\hfill  
\begin{subfigure}{0.48\textwidth}  
    \centering  
    \includegraphics[width=\linewidth]{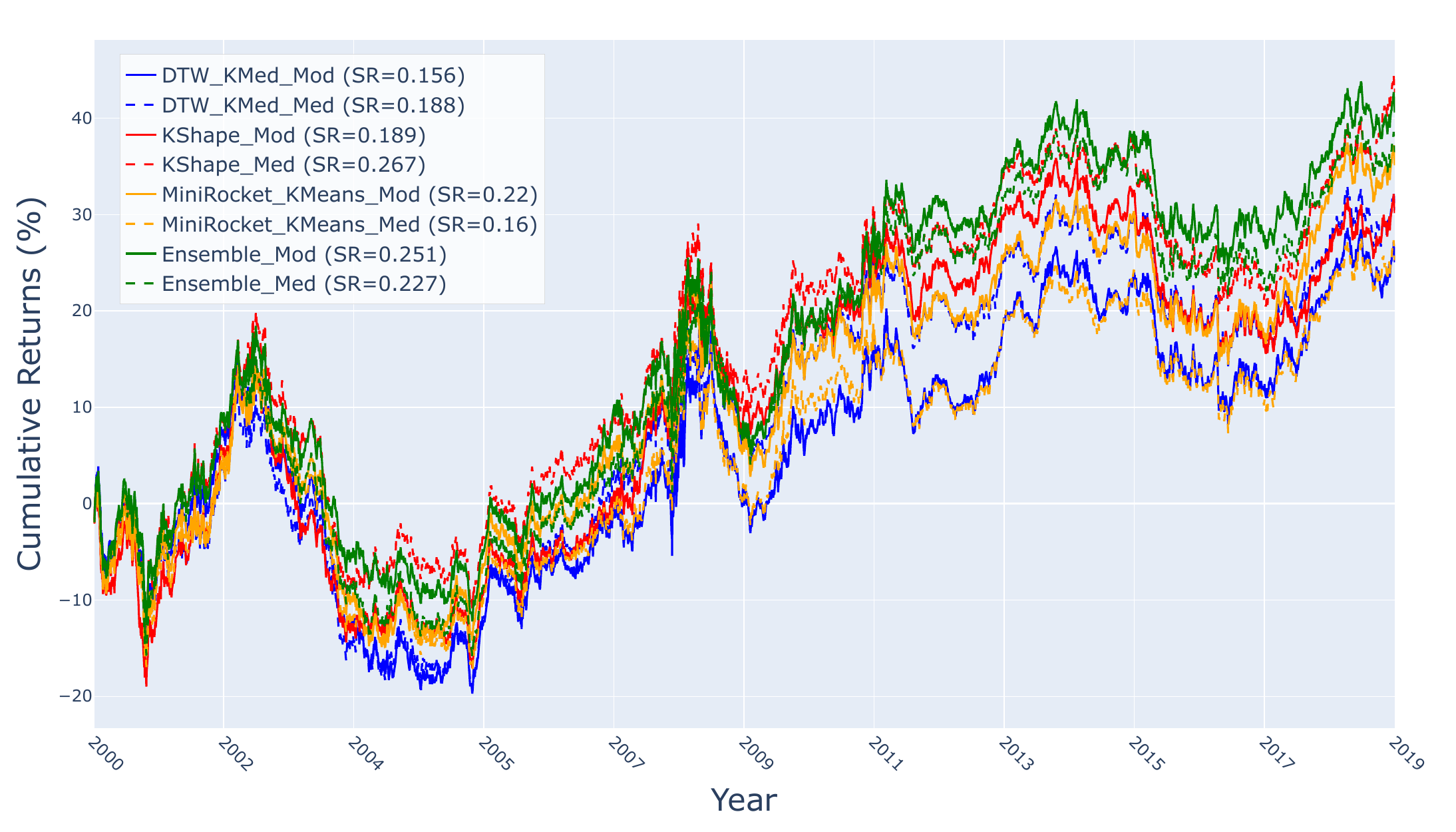}  
    \caption{Lead strategy on 1028 assets.} 
    \label{fig:Nation2}  
\end{subfigure}  
\caption{Cumulative return (PnL) plots under different clustering algorithms for the lead strategy.}  
\label{fig:pnl_lead_grid}  
\end{figure*}
\begin{figure*}[t]  
\centering  
\begin{subfigure}{0.48\textwidth}  
    \centering  
    \includegraphics[width=\linewidth]{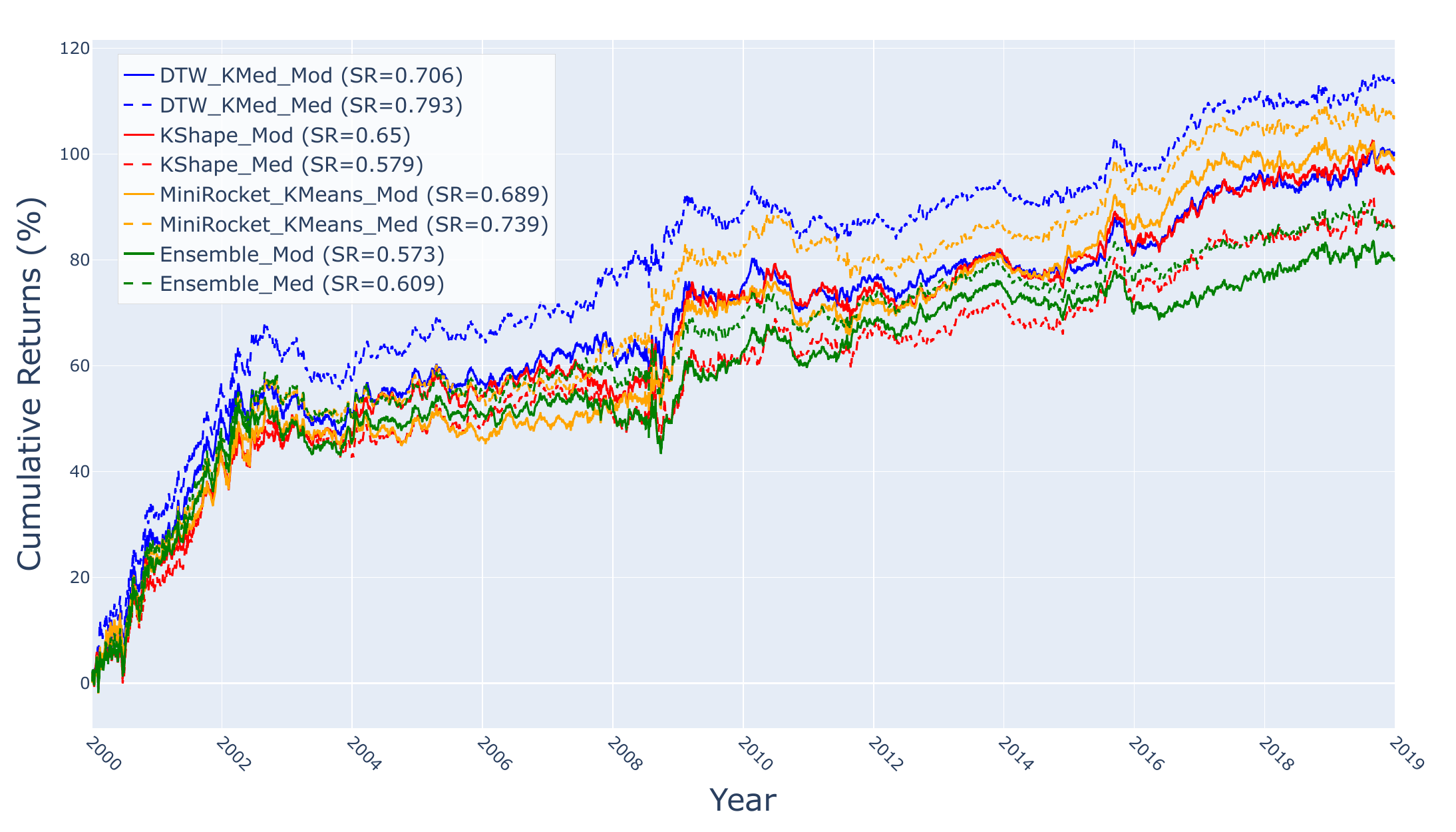}  
    \caption{Lag strategy on 679 assets.} 
    \label{fig:Nation1}  
\end{subfigure}  
\hfill  
\begin{subfigure}{0.48\textwidth}  
    \centering  
    \includegraphics[width=\linewidth]{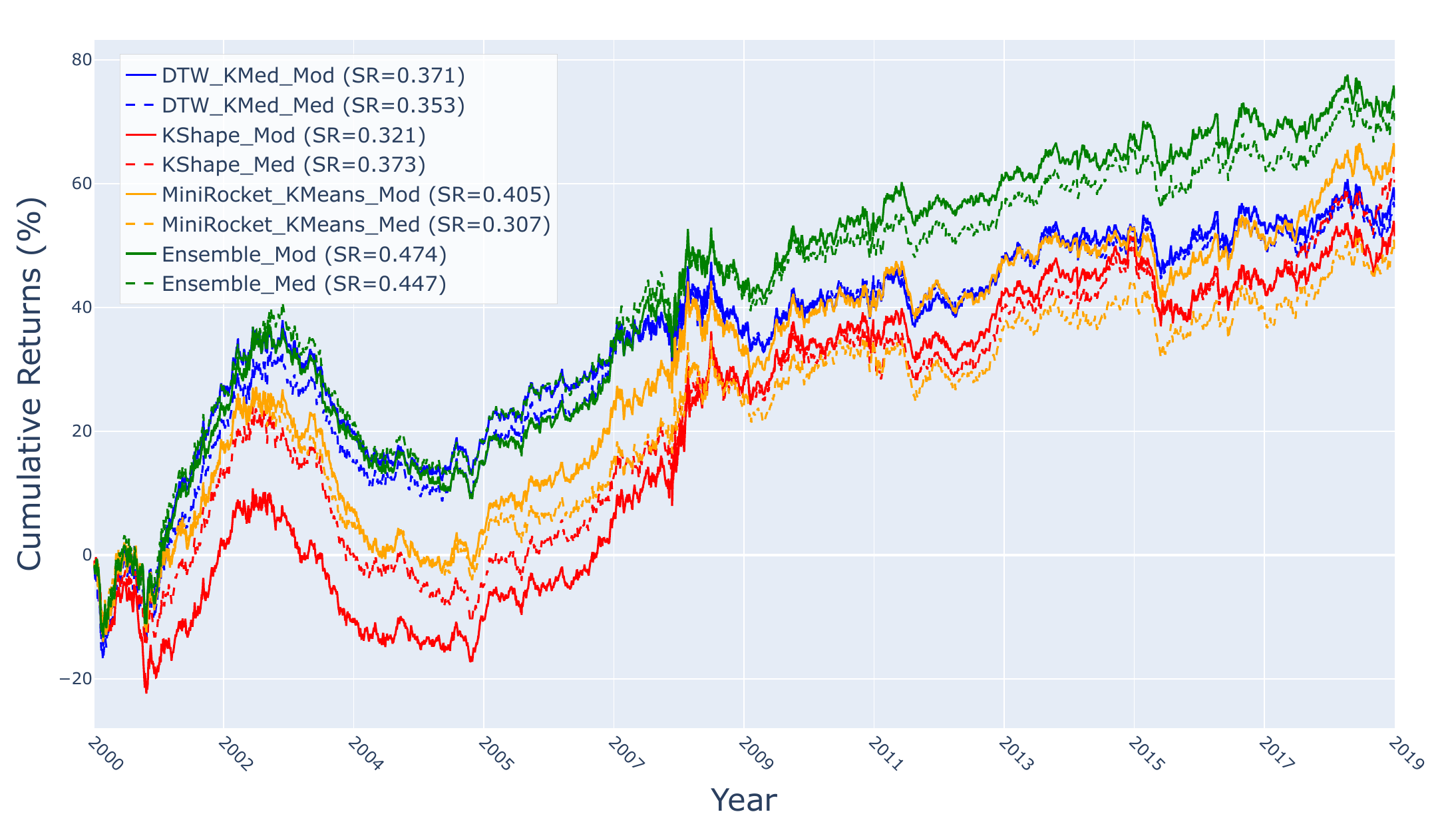}  
    \caption{Lag strategy on 1028 assets.} 
    \label{fig:Nation3}  
\end{subfigure}  
\caption{Cumulative return (PnL) plots under different clustering algorithms for the lag strategy.}  
\label{fig:pnl_lag_grid}  
\end{figure*}

The cumulative return plots of the lead strategy under different algorithms are shown in Figure~\ref{fig:pnl_lead_grid}, while the cumulative returns of the lag strategy are shown in Figure~\ref{fig:pnl_lag_grid}.

The plots clearly demonstrate the outstanding performance of the MiniRocket-KMeans algorithm and the robustness of the ensemble algorithm.





\begin{figure*}[t]  
\centering  
\begin{subfigure}{0.48\textwidth}  
    \centering  
    \includegraphics[width=\linewidth]{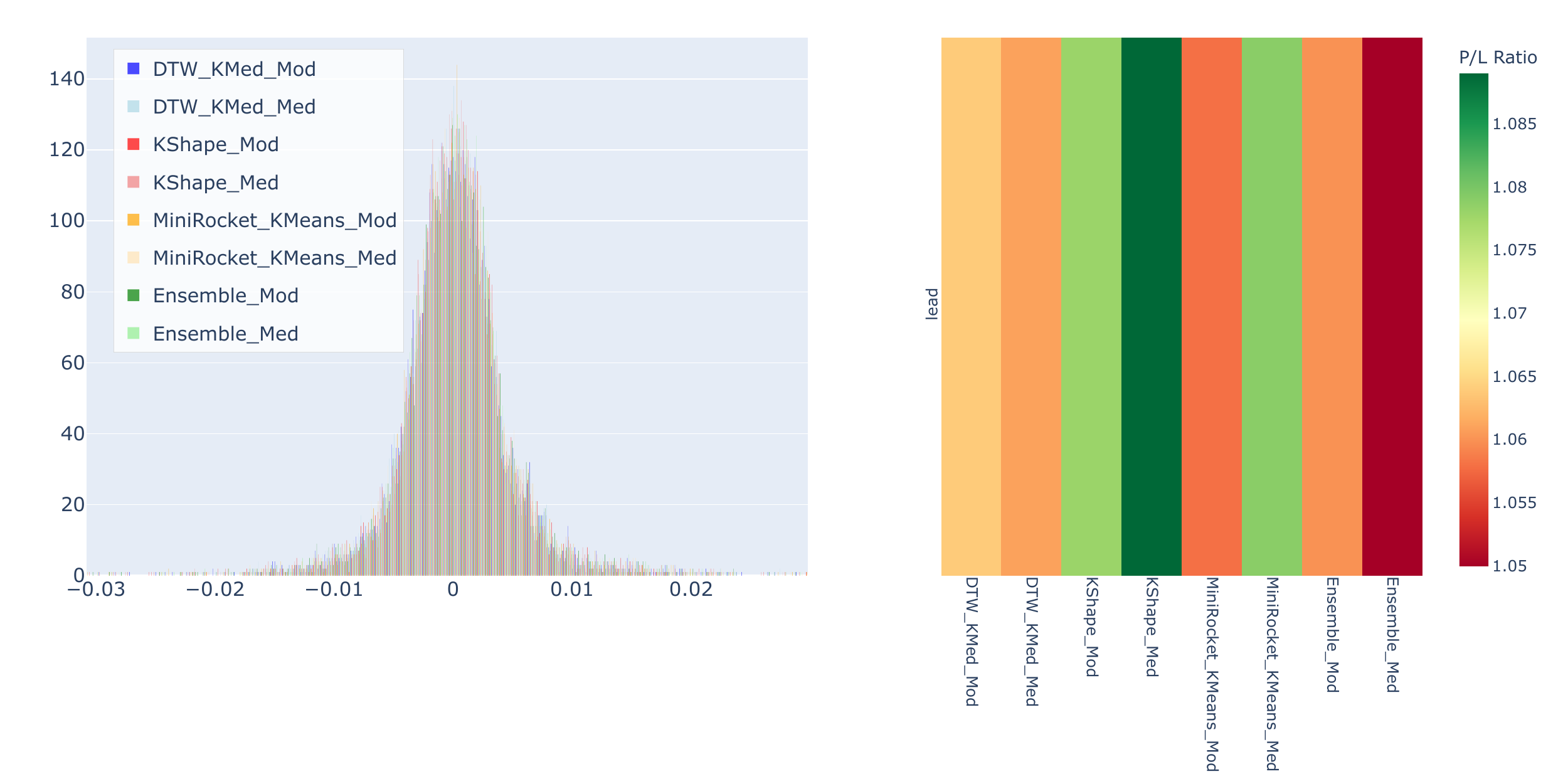}  
    \caption{Lead strategy on 679 assets.} 
    \label{fig:Nation6}  
\end{subfigure}  
\hfill  
\begin{subfigure}{0.48\textwidth}  
    \centering  
    \includegraphics[width=\linewidth]{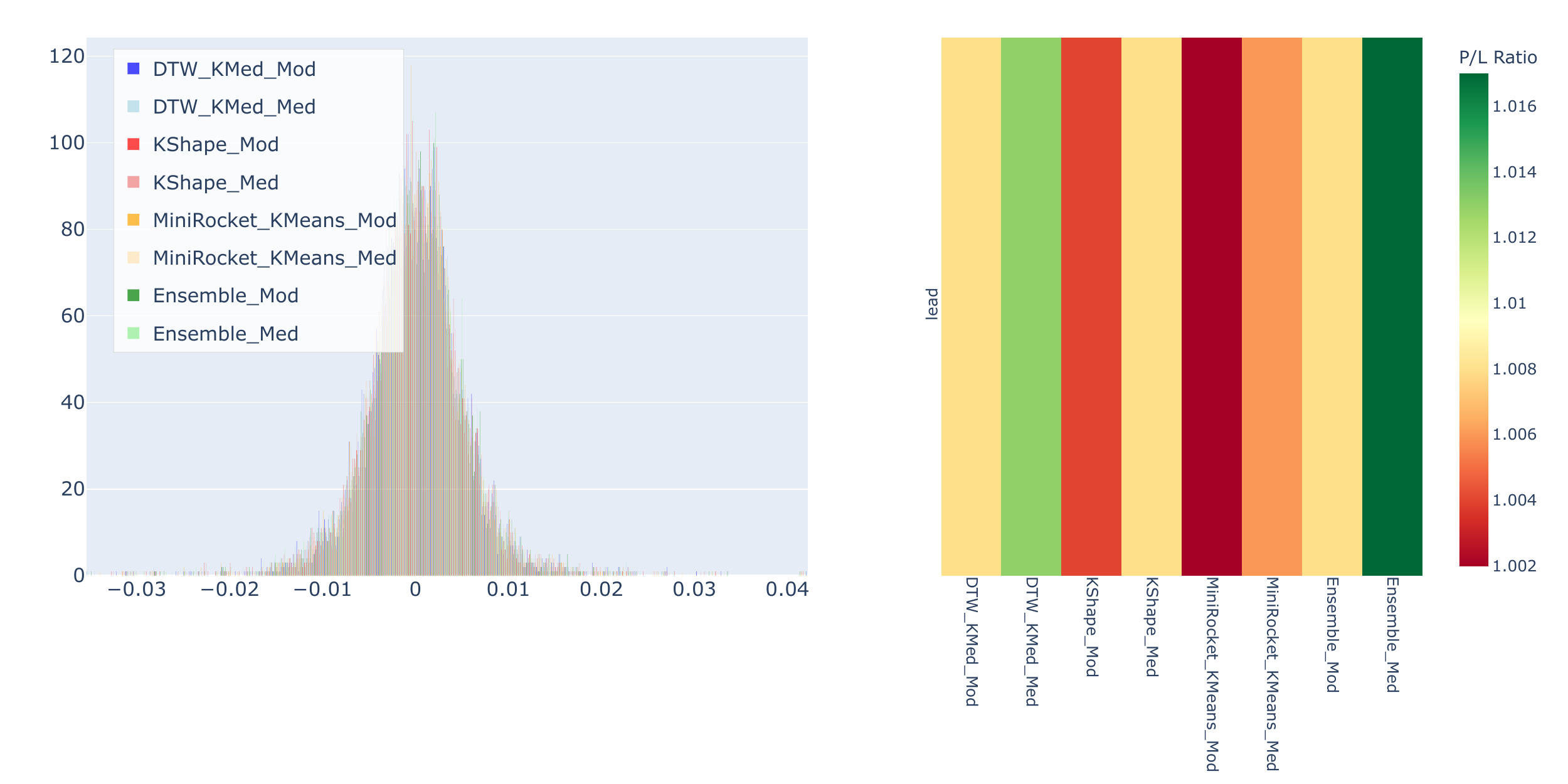}  
    \caption{Lead strategy on 1028 assets.} 
    \label{fig:Nation4}  
\end{subfigure}  
\caption{PnL distribution histograms (left plots within subfigures) and profit-loss ratio heatmaps (right plots within subfigures) under different clustering algorithms for the lead strategy.}  
\label{fig:pnl_dist_lead_grid}  
\end{figure*}
\begin{figure*}[t]  
\centering  
\begin{subfigure}{0.48\textwidth}  
    \centering  
    \includegraphics[width=\linewidth]{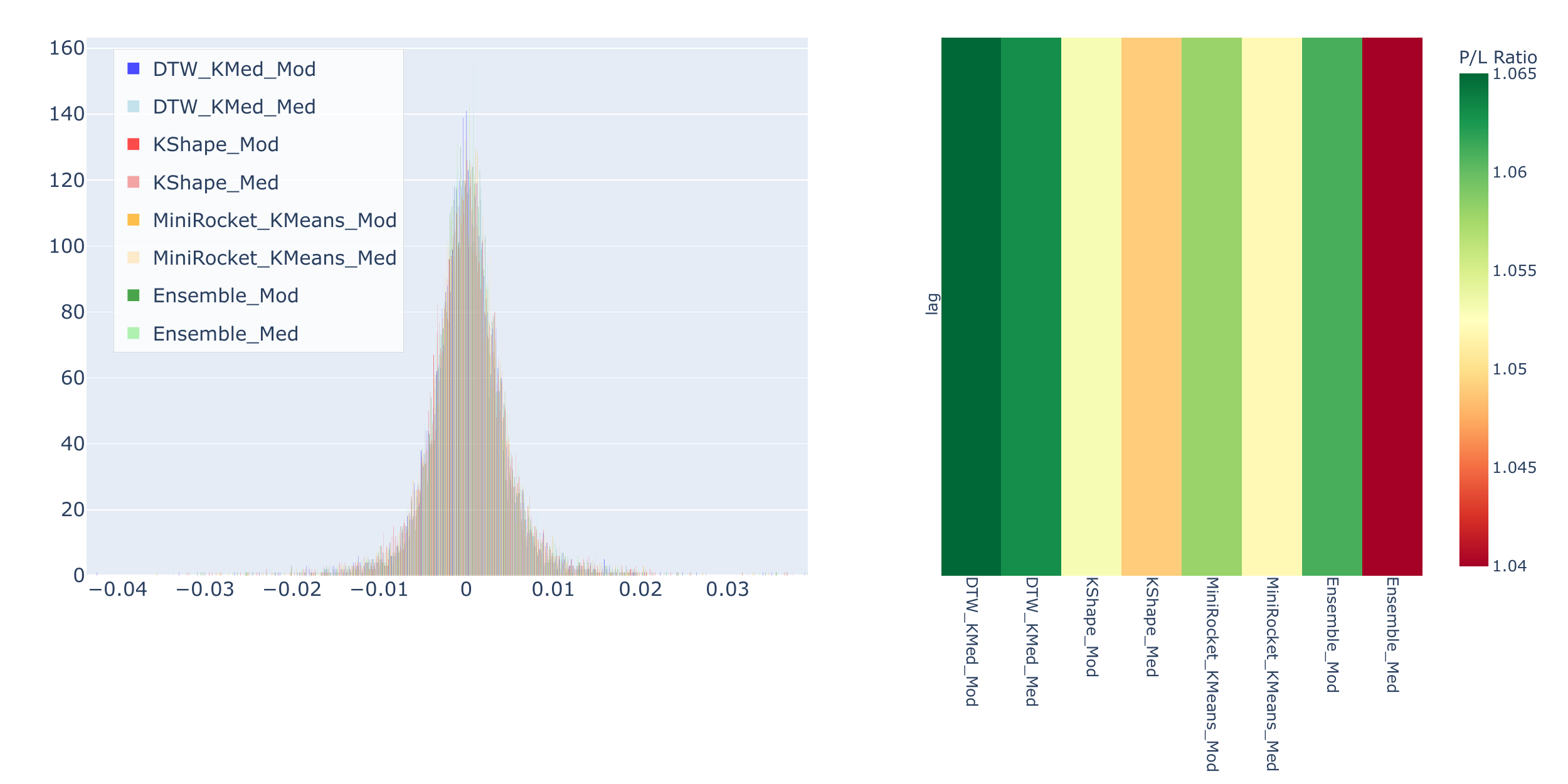}  
    \caption{Lag strategy on 679 assets.} 
    \label{fig:Nation7}  
\end{subfigure}  
\hfill  
\begin{subfigure}{0.48\textwidth}  
    \centering  
    \includegraphics[width=\linewidth]{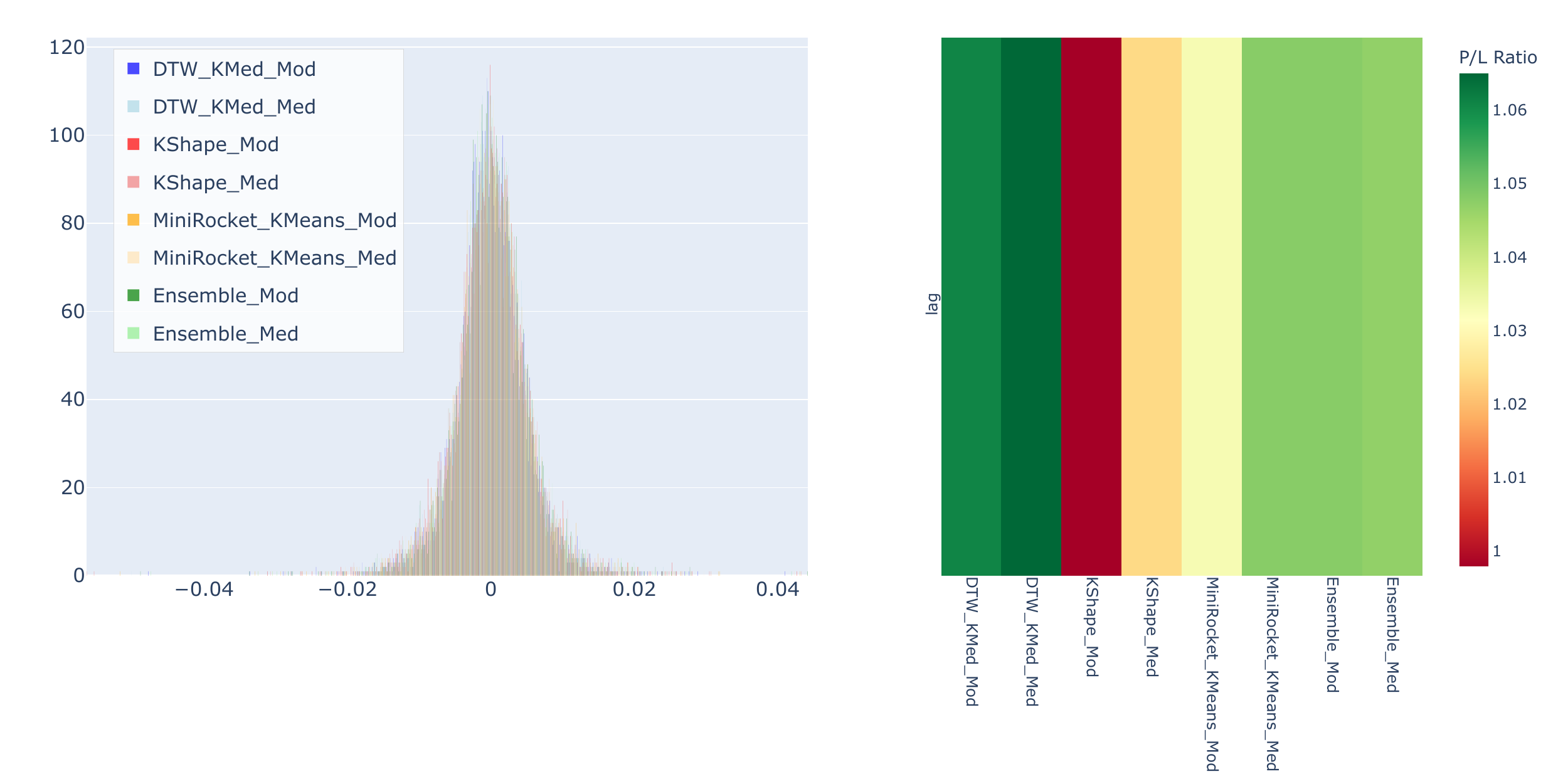}  
    \caption{Lag strategy on 1028 assets.} 
    \label{fig:Nation5}  
\end{subfigure}  
\caption{PnL distribution histograms (left plots within subfigures) and profit-loss ratio heatmaps (right plots within subfigures) under different clustering algorithms for the lag strategy.}  
\label{fig:pnl_dist_lag_grid}  
\end{figure*}

The PnL distribution histograms and profit-loss ratio heatmaps of the lead strategy under different algorithms are shown in Figure~\ref{fig:pnl_dist_lead_grid}, while those of the lag strategy are shown in Figure~\ref{fig:pnl_dist_lag_grid}.

It can be seen that the daily return distributions of all algorithms are similar: returns fluctuate around the mean, exhibit an approximately normal shape and contain a small number of extreme values. Such a distribution is reasonable. From the profit‑loss ratio heatmaps, the profit‑loss ratio of all algorithms is slightly greater than 1, which provides the foundation for positive expected returns of the strategies.

\section{Conclusion and Future Work}

Based on the widely used DTW-KMedoids algorithm, this paper introduces an algorithm to find the optimal number of clusters, which improves the robustness and clustering accuracy of the original algorithm. Meanwhile, other clustering approaches are adopted and evaluated.

Future improvements mainly include: First, customizing the lead-lag matrix calculation framework according to specific clustering properties instead of using a unified DTW alignment. Second, exploring combinations of MiniRocket feature extraction with other robust clustering approaches in Euclidean space (such as HDBSCAN). Third, extending the hard voting ensemble model with a soft voting mechanism based on real-time silhouette coefficient weights. Fourth, optimizing cluster coherence to address the triangle inequality limitation of the DTW distance.

\bibliographystyle{ACM-Reference-Format}
\bibliography{reference}

@article{albers2021fragmentation,
  title   = {Fragmentation, Price Formation and Cross-Impact in BitcoinMarkets},
  author  = {Jakob Albers and Mihai Cucuringu and Sam Howison \& Alexander Y. Shestopaloﬀ},
  journal = {Applied Mathematical Finance},
  year    = {2021},
  pages   ={395-448},
  doi     = {10.1080/1350486X.2022.2080083}
}

@article{buccheri2021highfrequency,
  title   = {High-frequency lead-lag effects and cross-asset linkages: a multi-asset lagged adjustment model},
  author  = {G. Buccheri and F. Corsi and S. Peluso},
  journal = {Journal of Business \& Economic Statistics},
  year    = {2021},
  publisher = {Taylor \& Francis},
  doi     = {10.1080/07350015.2019.1697699}
}

@article{ito2020direct,
  title   = {Direct Estimation of Lead-Lag Relationships Using Multinomial Dynamic Time Warping},
  author  = {Ito and K. and Sakemoto and R. },
  journal = {Asia-Pacific Financial Markets},
  volume  = {27},
  number  = {3},
  pages   = {325--342},
  year    = {2020},
  publisher = {Springer},
  doi     = {10.1007/s10690-019-09295-z}
}

@article{DTW_leadlag_lagged,
  title   = {Dynamic Time Warping for Lead-Lag Relationships in Lagged Multi-Factor Models},
  author  = {Yichi Zhang and Mihai Cucuringu and Alexander Y. Shestopaloff and Stefan Zohren},
  journal = {arXiv preprint arXiv:2309.08800},
  year    = {2023},
  url     = {https://doi.org/10.48550/arXiv.2309.08800}
}

@article{zhang2023robust,
  title   = {Robust Detection of Lead-Lag Relationships in Lagged Multi-Factor Models},
  author  = {Yichi Zhang and Mihai Cucuringu and Alexander Y. Shestopaloff and Stefan Zohren},
  journal = {arXiv:2305.06704},
  year    = {2023},
  url     = {https://doi.org/10.48550/arXiv.2305.06704}
}

@inproceedings{multiref_leadlag,
  title     = {Multireference Alignment for Lead-Lag Detection in Multivariate Time Series and Equity Trading},
  author    = {Danni Shi and Jan-Peter Calliess and Mihai Cucuringu},
  booktitle = {Proceedings of the Fourth ACM International Conference on AI in Finance},
  pages     = {507--515},
  year      = {2023},
  publisher = {ACM},
  doi       = {10.1145/3604237.3626879}
}

@article{bellman1959adaptive,
  author={R. Bellman and R. Kalaba},
  journal={IRE Transactions on Automatic Control}, 
  title={On adaptive control processes}, 
  year={1959},
  volume={4},
  number={2},
  pages={1-9},
  doi={10.1109/TAC.1959.1104847}}

@inproceedings{berndt1994dynamic,
  title     = {Using Dynamic Time Warping to Find Patterns in Time Series},
  author    = {Donald J. Berndt and James Clifford},
  booktitle = { Proceedings of the 3rd International Conference on Knowledge Discovery and Data Mining},
  pages     = {359--370},
  year      = {1994},
  url       = {https://dl.acm.org/doi/abs/10.5555/3000850.3000887}
}

@article{keogh2005exact,
  title   = {Exact Indexing of Dynamic Time Warping},
  author  = {Eamonn Keogh and Chotirat Ann Ratanamahatana},
  journal = {Knowledge and Information Systems},
  volume  = {7},
  pages   = {358--386},
  year    = {2005},
  publisher = {Springer},
  doi     = {10.1007/s10115-004-0154-9}
}

@techreport{DTW_review,
  title       = {Dynamic Time Warping Algorithm Review},
  author      = {Pavel Senin},
  institution = {University of Hawaii at Manoa, Department of Information and Computer Science},
  pages       = {1--23},
  year        = {2008},
  month       = {December},
  address     = {Honolulu, USA},
  url         = {https://www.researchgate.net/profile/Pavel-Senin/publication/228785661_Dynamic_Time_Warping_Algorithm_Review/links/02bfe5100f11a7929f000000/Dynamic-Time-Warping-Algorithm-Review.pdf}
}

@inproceedings{minirocket,
  title     = {{MiniRocket}: A Very Fast (Almost) Deterministic Transform for Time Series Classification},
  author    = {Angus Dempster and Daniel F. Schmidt and Geoffrey I. Webb},
  booktitle = {Proceedings of the 27th ACM SIGKDD Conference on Knowledge Discovery and Data Mining},
  pages     = {248--257},
  year      = {2021},
  publisher = {ACM},
  doi       = {10.1145/3447548.3467231}
}

@inproceedings{KShape,
  title     = {{k-Shape}: Efficient and Accurate Clustering of Time Series},
  author    = {John Paparrizos and Luis Gravano},
  booktitle = {Proceedings of the 2015 ACM SIGMOD International Conference on Management of Data},
  pages     = {1855--1870},
  year      = {2015},
  publisher = {ACM},
  doi       = {10.1145/2723372.2737793}
}

@article{jegadeesh2001profitability,
  title   = {Profitability of Momentum Strategies: An Evaluation of Alternative Explanations},
  author  = {Narasimhan Jegadeesh and Sheridan Titman},
  journal = {The Journal of Finance},
  volume  = {56},
  pages   = {699--720},
  year    = {2001},
  publisher = {Wiley},
  doi     = {10.1111/0022-1082.00342}
}

@article{lim2019enhancing,
  title   = {Enhancing Time-Series Momentum Strategies Using Deep Neural Networks},
  author  = {Bryan Lim and Stefan Zohren and Stephen Roberts},
  journal ={arXive preprint arXiv.1904.04912},
  year    = {2019},
  doi     ={10.48550/arXiv.1904.04912}
}

@article{poh2022transfer,
  title   = {Transfer Ranking in Finance: Applications to Cross-Sectional Momentum with Data Scarcity},
  author  = {Daniel Poh and Stephen Roberts and Stefan Zohren},
  journal = {arXiv preprint arXiv:2208.09968},
  year    = {2022},
  url     = {https://arxiv.org/abs/2208.09968}
}

@article{tan2023spatiotemporal,
  title   = {Spatio-Temporal Momentum: Jointly Learning Time-Series and Cross-Sectional Strategies},
  author  = {Wee Ling Tan and Stephen Roberts and Stefan Zohren},
  journal = {arXiv preprint arXiv:2302.10175},
  year    = {2023},
  url     = {https://arxiv.org/abs/2302.10175}
}

@article{wood2021trading,
  title   = {Trading with the Momentum Transformer: An Intelligent and Interpretable Architecture},
  author  = {Kieran W. and Sven G.and Stephen R. and Stefan Z.},
  journal = {arXiv preprint arXiv:2112.08534},
  year    = {2021},
  url     = {https://arxiv.org/abs/2112.08534}
}

@article{wood2021slow,
  title   = {Slow Momentum with Fast Reversion: A Trading Strategy Using Deep Learning and Changepoint Detection},
  author  = {Kieran Wood and Stephen Roberts and Stefan Zohren},
  journal = {arXiv preprint arXiv:2105.13727},
  year    = {2021},
  url     = {https://arxiv.org/abs/2105.13727}
}

@article{roberts2023network,
  title   = {Network Momentum across Asset Classes},
  author  = {Xingyue P. and Stephen R. and Xiaowen D.and Stefan Z.},
  journal = {arXiv preprint arXiv:2308.11294},
  year    = {2023},
  url     = {https://arxiv.org/abs/2308.11294}
}

@article{zohren2023learning,
  title   = {Learning to Learn Financial Networks for Optimising Momentum Strategies},
  author  = {Xingyue Pu and Stefan Zohren and Stephen Roberts and Xiaowen Dong},
  journal = {arXiv preprint arXiv:2308.12212},
  year    = {2023},
  url     = {https://arxiv.org/abs/2308.12212}
}

\end{document}